\documentclass[prl,twocolumn]{revtex4}

\pdfoutput=1
\usepackage{graphics}
\usepackage{graphicx}   

\begin{document}

\title{Correlated volume-energy fluctuations of phospholipid membranes: \newline A simulation study}

\author{Ulf R. Pedersen}
\affiliation{DNRF Centre ``Glass and Time,'' IMFUFA, Department of Sciences, Roskilde University, Postbox 260, DK-4000 Roskilde, Denmark.}
\affiliation{Department of Chemistry, University of California, Berkeley, California 94720-1460, USA}
\email{urp@berkeley.edu}
\author{G\"{u}nther H. Peters}
\affiliation{Center for Biopmembrane Physics (MEMPHYS), Department of Chemistry, Technical University of Denmark, DK-2800 Kgs. Lyngby, Denmark}
\author{Thomas B. Schr{\o}der}
\affiliation{DNRF Centre ``Glass and Time,'' IMFUFA, Department of Sciences, Roskilde University, Postbox 260, DK-4000 Roskilde, Denmark.}
\author{Jeppe C. Dyre}
\affiliation{DNRF Centre ``Glass and Time,'' IMFUFA, Department of Sciences, Roskilde University, Postbox 260, DK-4000 Roskilde, Denmark.}

\begin{abstract}
This paper reports all-atom computer simulations of five phospholipid membranes (DMPC, DPPC, DMPG, DMPS, and DMPSH) with focus on the thermal equilibrium fluctuations of volume, energy, area, thickness, and chain order. At constant temperature and pressure, volume and energy exhibit strong correlations of their slow fluctuations (defined by averaging over 0.5 nanosecond). These quantities, on the other hand, do not correlate significantly with area, thickness, or chain order. The correlations are mainly reported for the fluid phase, but we also give some results for the ordered (gel) phase of two membranes, showing a similar picture.
The cause of the observed strong correlations is identified by splitting volume and energy into contributions from tails, heads, and water, and showing that the slow volume-energy fluctuations derive from van der Waals interactions of the tail region; they are thus analogous to the similar strong correlations recently observed in computer simulations of the Lennard-Jones and other simple van der Waals type liquids [U. R. Pedersen et al., Phys. Rev. Lett. 2008, 100, 015701].
The strong correlations reported here confirm one crucial assumption of a recent theory for nerve signal propagation proposed by Heimburg and Jackson [T. Heimburg and A. D. Jackson, Proc. Natl. Acad. Sci. 2005, 102, 9790-9795].
\end{abstract}

\date{\today}

\maketitle

In 2005 Heimburg and Jackson showed that biomembranes may carry solitonic sound waves whose maximum amplitude and minimum velocity are close to the propagation velocity in myelinated nerves \cite{Heimburg2005}. Their paper concluded: ``It would be surprising if nature did not exploit these features.'' Subsequent works by the same authors argue directly that nerve signals are not primarily electrical, but solitonic sound waves carried by the nerve cell membrane \cite{Heimburg2007A,Andersen2009}. The conventional wisdom is that nerve signals propagate via electrical current as formulated in the Hodgkin-Huxley theory \cite{Hodgkin1952}. The Heimburg-Jackson theory explains anaesthesia as a straightforward effect of melting-point depression of the order-disorder transition. This can in turn explain why biomembranes have a transition temperatures close to physiological temperatures \cite{Nielsen2007}.

The Heimburg-Jackson nerve-signal theory motivated this study. One element of the theory is the assumption that volume and enthalpy correlate strongly in their thermal equilibrium fluctuations, at least as it regards the slow parts of these fluctuations. This assumption was justified by experimental observations that membranes specific heat and compressibility are proportional in their temperature dependence across the phase transition \cite{Heimburg1998,Ebel2001}. Also, a correlation between lipid area and enthalpy has been proposed \cite{Heimburg1998} based on experimental observations\cite{dimova2000}. If so, both volume, membrane area and enthalpy are controlled by a single parameter. Below, we present first results from extensive computer simulations of different phospholipid membranes performed in order to investigate whether strong volume-energy and/or area-energy correlations are observed and, if so, what causes them.

Recently, we studied equilibrium thermodynamic fluctuations of much simpler systems, namely various model liquids like the standard Lennard-Jones (LJ) liquid and similar systems \cite{Pedersen2008PhysRev,Pedersen2008PhysRevE,Bailey2008A,Bailey2008B,bailey2008_jpcm}. In many such simple liquids one finds a strong correlation between the equilibrium fluctuations of virial $W$ and potential energy $U$, when fluctuations are studied at constant particle number $N$, constant volume $V$, and constant temperature $T$ (the so-called $NVT$ ensemble \cite{Feller1995}). Recall\cite{Allen1989}, that the virial $W=W(t)$ gives the non-ideal contribution to the instantaneous pressure $p=p(t)$ via the defining equation: $p(t)V=Nk_BT(t)+W(t)$ where $T(t)$ is the instantaneous temperature defined via the instantaneous kinetic energy. The virial is a function of the particle positions \cite{Allen1989}. For the LJ liquid, as well as for a united-atom toluene model, a dumbbell model, the Kob-Andersen binary LJ liquid \cite{Kob1994}, and other van der Waals liquids, $W$ and $U$ correlate better than 90\% in their equilibrium fluctuations \cite{Pedersen2008PhysRev}. This reflects an effective inverse power-law potential dominating fluctuations, as discussed in details in Refs. \onlinecite{Bailey2008A,Bailey2008B}. Liquids with poor $W-U$ correlation include water and methanol \cite{Pedersen2008PhysRev,Bailey2008A}. In these cases the correlations are ruined by the hydrogen bonds, which are conventionally modelled via Coulomb forces -- the existence of competing interactions prevents strong $W-U$ correlation in hydrogen-bonded liquids.
%\citenum{Bailey2008A,Bailey2008B}

For liquids with time-scale separation, like highly viscous liquids, strong $W-U$ correlations are particularly significant: Viscous liquids with strong $W-U$ correlations are close to being "single-order-parameter liquids" in the classical Prigogine-Defay \cite{prigogine1954,Davies1952} sense\cite{Pedersen2008PhysRevE,Ellegaard2007,bailey2008_jpcm}. This implies that complex frequency-dependent thermoviscoelastic response functions like the isobaric/isochoric dynamic specific heat, the dynamic thermal expansion coefficient, the dynamic compressibility, etc, are all given by a single function \cite{Ellegaard2007}. In particular, these cannot ``decouple'' \cite{Angell1991} from one another -- they must all exhibit relaxations in the same frequency range. It has also been shown that strongly correlating viscous liquids obey density scaling, i.e., that if the relaxation time $\tau$ is measured at different temperatures and density $\rho$, $\tau$ is a unique function \cite{Roland2005} of $\rho^\gamma/T$ where the exponent $\gamma$ may be determined from studies of $W-U$ correlations at a single state point \cite{Schroder2008}. Finally, it was recently found that strongly correlating viscous liquids have much simpler aging properties than viscous liquids in general \cite{schroder2009}.

Fluctuations are ensemble dependent, of course, and one may ask what happens if fluctuations are studied instead in the ensemble of constant temperature and pressure ($NpT$ ensemble). In this ensemble virial fluctuations are not interesting, but there are strong correlations for simple liquids between the fluctuations of volume and potential energy \cite{Pedersen2008PhysRevE}. This is the ensemble used below for studying biomembrane thermodynamic fluctuations. 

Motivated by the findings for simple model liquids and the Heimburg-Jackson theory, we decided to focus on phospholipid membranes in order to investigate whether the Heimburg-Jackson assumption of strong volume-enthalpy correlations is confirmed. A phospholipid has van der Waals interactions between its acyl chains and hydrogen bonds in the head region. Similarities to simple van der Waals liquids are not at all obvious, and the microscopic origin of the volume-enthalpy correlation tentatively derived from experiments \cite{Heimburg1998,Ebel2001} is not trivial.
Thermodynamic fluctuations of simulated membranes have been studied in the past \cite{jeu2003,bolterauer1991,feller1996,baron2006}, however, not with a focus on volume-energy correlations.

Because of the large amount of water in the system and the hydrophilic head groups, one does not expect strong instantaneous correlations of phospholipid membrane thermodynamic fluctuations \cite{Pedersen2008PhysRev,Bailey2008A}. The Heimburg-Jackson theory, however, relates to strong correlations of slow degrees of freedom of the biomembrane (on millisecond time scales), and so do the experiments they quote indicating strong correlations \cite{Heimburg1998,Ebel2001}. This is analogous to the situation for highly viscous liquids where time-scale separation between the fast, vibrational degrees of freedom and the much slower configurational ones is also e.g. essential for viscous flow or visco-elastic responses.
Phospholipids are the major constituent of biological membranes \cite{heimburg2007book}. Close to physiological temperature these membranes undergo a transition from an ordered phase (L$_\beta$) to a disordered phase (L$_\alpha$) \cite{jeu2003,nagle1980,Nielsen2007,Zhang1995,jin1997}. Below, we evaluate the strength of $V-U$ correlations for both the disordered and ordered phases with main focus on the disordered phase. The correlation strengths are calculated for a range of time scales. We show that $V-U$ fluctuations correlate strongly, but only on long time scales \cite{pedersen2008aip}. We also investigate how well membrane area as well as chain order-parameter fluctuations correlate with $V$ and $U$ fluctuations; such correlations are generally weak. Finally, the cause of the correlations is identified by splitting volume and energy into contributions from the tail-, the head-, and the water region: The slow, strongly correlating $V-U$ fluctuations are shown to derive from the tails which are dominated by van der Waals interactions, thus establishing a conceptual link to the strong correlations of the slow pressure-energy fluctuations -- the virial/potential energy correlations -- of simple van der Waals liquids \cite{Pedersen2008PhysRev}.

\section{Simulation details}

\begin{figure}
\begin{center}
\includegraphics[width=0.24\textwidth]{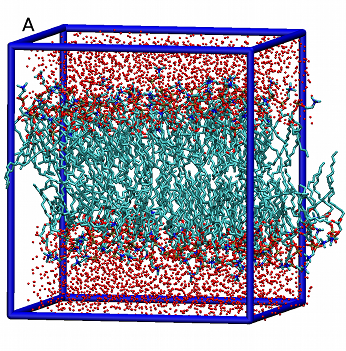}
\includegraphics[width=0.23\textwidth]{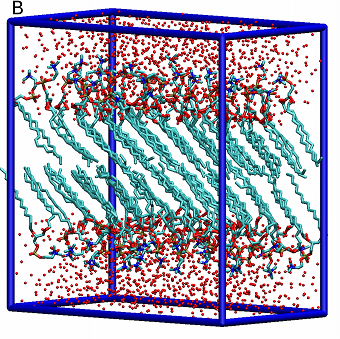}
\end{center}
\caption{Snapshots of DMPC membranes in the fluid phase (DMPC-f; panel A) and in the ordered phase (DMPC-g; panel B). The red atoms are the oxygen atoms of water molecules; hydrogen atoms were removed for visual clarity (but included in the simulations). Acyl chains are colored green. The frame indicates the periodic boundary box.}\label{fluidNgel}
\end{figure}

\begin{table*}
\caption{Overview of simulation details and results}\label{Table1}

\begin{tabular}{l|cccccc}
 (A) System (phase) &  $t_\textrm{sim}$ [ns] & $t_\textrm{prod}$ [ns] & $N_\textrm{lipid}$ & $T$ [K] & $T_\textrm{actual}$ [K] & $N_\textrm{water}/N_\textrm{lipid}$ \\
\hline
DMPC-f (fluid) & 151 & 121 & 128 & 330.0 & 329.0$\pm$1.6 & 33  \\
%DMPC-neq (non eq.) & 167 & (0.969 & 0.884 & 0.883)$^*$ & 6.33 \\
%DMPC-g1 (ordered) & 20  & 64  & 286 & 11 & 0.77 & -0.35 & 0.07 & 3.1 \\
DMPC-g (ordered) & 65 & 36  & 64  & 286.0 & 285.3$\pm$2.0 & 33 \\
\hline
DPPC-f (fluid) & 180 & 124 & 72  & 325.0 & 324.0$\pm$2.1  & 29  \\
%DPPC-g1 (ordered)  & 68 & 64  & 304 & 11 & 0.85 & 0.02 & 0.09 & 3.6  \\
DPPC-g (ordered) & 78 & 48 & 64 & 304.0 & 303.2$\pm$2.1 & 33  \\
%DPPC-g2b (ordered)& 17 & 64 & 294 & 33 & 0.906 & -0.119 & 0.087 & 4.00 \\
\hline
%(DMPG$^1$      89 & 0.933 & 0.701 & 0.627  & 5.84) \\
DMPG (fluid) & 149 & 49 & 128  & 330.0 & 329.0$\pm$1.6 & 33   \\
DMPS (fluid) & 139 & 49  & 128  & 340.0 & 339.0$\pm$1.7 & 36   \\
DMPSH (fluid) & 136 & 35  & 128  & 340.0 & 339.1$\pm$1.6 & 37  \\
\end{tabular}

\vspace{0.2cm}

\begin{tabular}{l|cc|c|ccccccc}
 (B) System (phase) & $R_{\bar{U}\bar{V}}$ & $\gamma$ & $R_{\bar{U}\bar{U}_{t}}$ & $R_{\bar{U}\bar{A}}$& $R_{\bar{V}\bar{A}}$ & $R_{\bar{A}\bar{S}_{CD}}$ & $R_{\bar{U}\bar{S}_{CD}}$ & $R_{\bar{V}\bar{S}_{CD}}$ \\
\hline
DMPC-f (fluid) & 0.77$\pm0.04$ &   9.6 & 0.82 & 0.50 & 0.57 & -0.75 & -0.49 & -0.54 \\
DMPC-g (ordered) & 0.47$\pm0.07$ & 6.1 & 0.31 & 0.02 & 0.05 & -0.64 & 0.12 & 0.14  \\
\hline
DPPC-f (fluid) & 0.87$\pm0.04$ & 10.2 & 0.89 & -0.29 & -0.36 & 0.00 & -0.61 & -0.71 \\
DPPC-g (ordered) &  0.75$\pm0.07$ & 6.6 & 0.71 & -0.16 & 0.12 & -0.67 & 0.09 & -0.07 \\
\hline
DMPG (fluid) & 0.82$\pm0.07$ & 8.5 &0.80 & 0.41 & 0.40 & -0.76 & 0.01 & 0.08 \\
DMPS (fluid) & 0.59$\pm0.07$ & 7.6 &0.64 & 0.30 & 0.28 & -0.71 & 0.04 & 0.20 \\
DMPSH (fluid) & 0.78$\pm0.07$ & 13.2 &0.84 & 0.43 & 0.51 & -0.50 & 0.05 & 0.14 \\
\end{tabular}

\vspace{0.2cm}

(A) $t_\textrm{sim}$: Total simulation time in nanoseconds. $t_\textrm{prod}$: Length of production run in nanoseconds (only membranes in quasi-equilibrium, i.e., with no detectable drift in the area per molecule, were included in the data analysis); $N_\textrm{lipid}$: Number of lipid molecules; $T$: Thermostat temperature in Kelvin; $T_\textrm{actual}$: Average temperature and standart diviation of fluctuations; $N_\textrm{water}/N_\textrm{lipid}$: Number of water molecules per lipid molecule. 
(B) $R_{\bar{U}\bar{V}}$: Energy-Volume correlation coefficient (see Equation \ref{corrcoef}). The bar indicated a 0.5 ns average. The uncertainty is estimated from the DMPC-f and DPPC-f trajectories as described in the text (67\% confidence interval); $\gamma$: Energy-volume scaling factor in units of $10^{-4}$ mL/J (see Equation \ref{gammadef}); $R_{\bar{U}\bar{U}_t}$: Energy-``Energy of tail region'' correlation coefficient. $R_{\bar{U}\bar{A}}$: Energy-Area correlation coefficient; $R_{\bar{V}\bar{A}}$: Volume-Area correlation coefficient. $R_{\bar{A}\bar{S}_{CD}}$: Area-``chain order-parameter'' correlation coefficient; $R_{\bar{U}\bar{S}_{CD}}$: Energy-``chain order-parameter'' correlation coefficient; $R_{\bar{V}\bar{S}_{CD}}$: Volume-``chain order-parameter'' correlation coefficient.

\end{table*}

Details of the seven membrane simulations performed are listed in Table \ref{Table1}. The following abbreviations are used:  
DMPC-f and DMPC-g: fully hydrated di-myristoyl-phosphatidyl-choline membrane in the fluid and ordered phases, respectively \cite{Pedersen2006}. Figure \ref{fluidNgel}A show a configuration of the DMPC-f membrane.
DPPC-f and DPPC-g: fully hydrated di-palmitoyl-phosphatidyl-choline membrane in the fluid and ordered phases, respectively \cite{Sonne2007}.
DPPG: a fully hydrated di-palmitoyl-phosphatidyl-glycerol membrane in the fluid phase with calcium counter ions \cite{Pedersen2007}.
DPPS: a fully hydrated di-myristoyl-phosphatidyl-serine membrane in the fluid phase with calcium counter ions \cite{Pedersen2007}.
DMPSH: a fully hydrated and protonated di-myristoyl-phosphatidyl-serine membrane in the fluid phase \cite{Pedersen2007}.
Phospholipids (and counter ions) were modeled using the all-atom CHARMM27 force-field \cite{macKerell1998-charmm} with modified charges of the head group as described in detail elsewhere \cite{Pedersen2006}. Membranes were hydrated using explicit water represented by the TIP3 model \cite{jorgensen1983-TIP3}.

Simulations were carried out using the NAMD software package \cite{Phillips2005}. In all simulations a time step of 1.0 fs was used. Temperature and pressure were controlled by a Langevin thermostat (damping coefficient: 5 ps$^{-1}$) and a Nos{\'e}-Hoover Langevin barostat (anisotropic regulation; piston oscillation time: 100 fs; damping time: 50 fs) \cite{Feller1995}. Electrostatic interactions were evaluated using the Particle-Mesh-Ewald method \cite{Darden1993,Essmann1995} with grid spacing of about 1\AA{} updated every 4 fs. 
Lennard-Jones potentials were brought smoothly to zero by multiplying a switching function $S(r_{ij})$;
$S(r_{ij})=1$ for $r_{ij}<R_\textrm{on}$, 
$S(r_{ij})=(R_\textrm{off}^2-r_{ij}^2)^2(R_\textrm{off}^2+2r_{ij}^2-3R_\textrm{on}^2)/(R_\textrm{off}^2-R_\textrm{on}^2)^3$ for $R_\textrm{on}<r_{ij}<R_\textrm{off}$ and
$S(r_{ij})=0$ for $r_{ij}>R_\textrm{off}$
with $R_\textrm{on}=10$ \AA{} and $R_\textrm{off}=12$ \AA{}.
Periodic boundary conditions were applied in all three dimensions.

Initial configurations of DMPC-f, DPPC-f, DPPG, DPPS, DMPS and DMPSH are taken from Refs.\ \onlinecite{Pedersen2006,Pedersen2007,Sonne2007}. Initial configurations of the ordered membranes, DMPC-g and DPPC-g, were build from a membrane simulated by Venable and co-workers \cite{Venable2000}. Throughout the simulation the acyl chains remain in an ordered structure as shown in Figure \ref{fluidNgel}B \cite{jin1997}.
Thermal equilibrium was ensured by monitoring the membrane area; only trajectories with no drift (compared to the thermal fluctuations) were used in the data analysis. Lengths of equilibrium trajectories are listed in Table \ref{Table1}A in the column under $t_\textrm{prod}$. The importance of firmly establishing that the systems is in equilibrium should be emphasized, since an apparent strong correlation would appear if volume and energy relax from some (arbitrary) out-of-equilibrium state. Only small finite size effects are expected at the simulated system sizes \cite{vries2005}.
%\citenum{Pedersen2006,Pedersen2007,Sonne2007}

\section{Results}

The following collective quantities were evaluated every 0.5 ps: Potential energy $U$, simulation box volume $V=XYZ$ (where $X$, $Y$ and $Z$ are the box dimensions), projected membrane area $A=XY$, box thickness $Z$, and average chain order parameter $\langle S_{CD}\rangle_{ch}$. The latter quantity characterizes the overall order of acyl chains \cite{Tieleman1996} and is defined by $\langle S_{CD}\rangle_{ch} = | \left\langle \frac{3}{2} \cos^2(\theta_{CD} )-\frac{1}{2} \right\rangle_{ch}|$ where $\theta_{CD}$ is the angle between the membrane normal ($\vec{z}$) and the C-H bond of a given methylene group, and $\langle \ldots \rangle_{ch}$ denotes an average over all methylene groups in all chains.

\begin{figure}
\begin{center}
\includegraphics[width=0.45\textwidth]{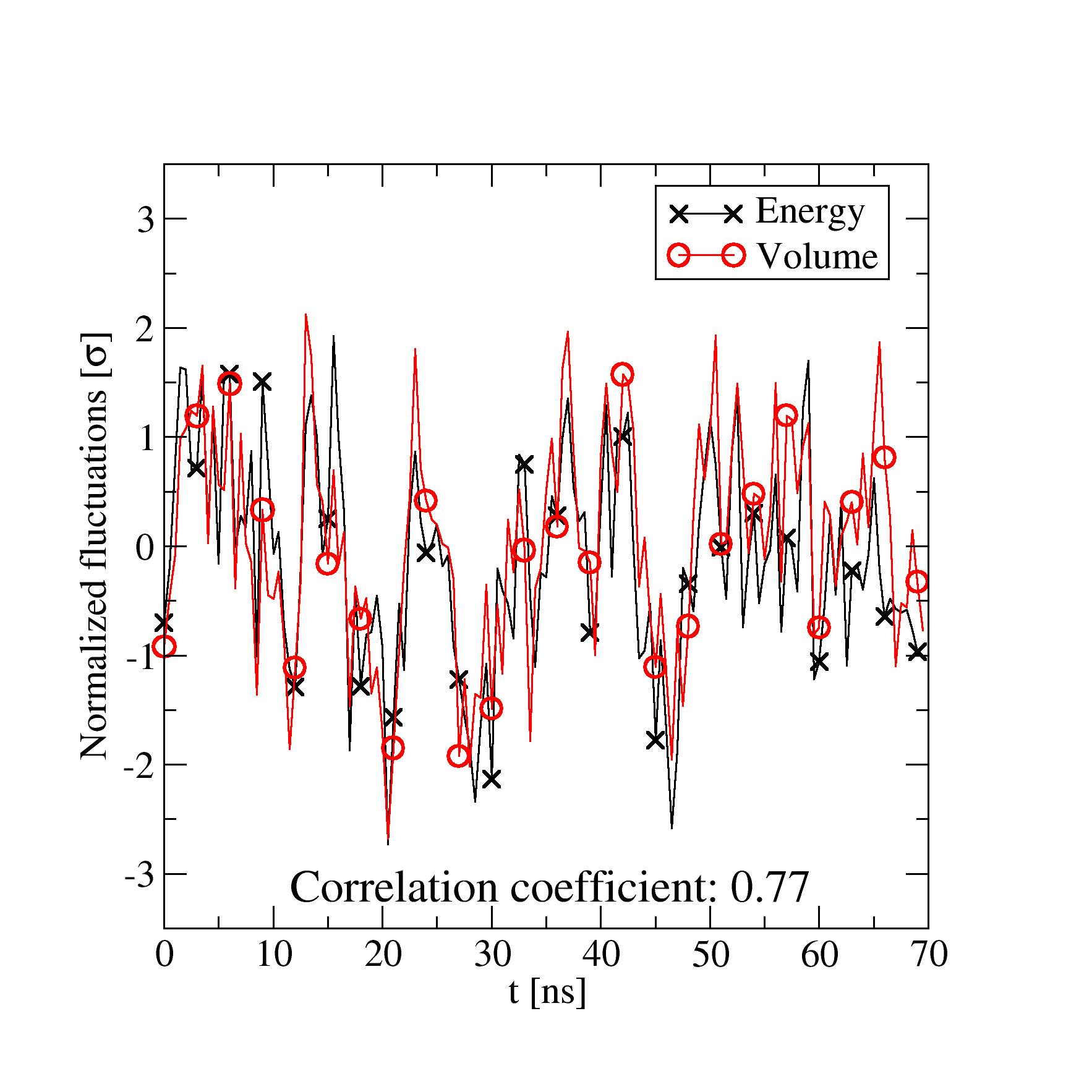}
\includegraphics[width=0.35\textwidth]{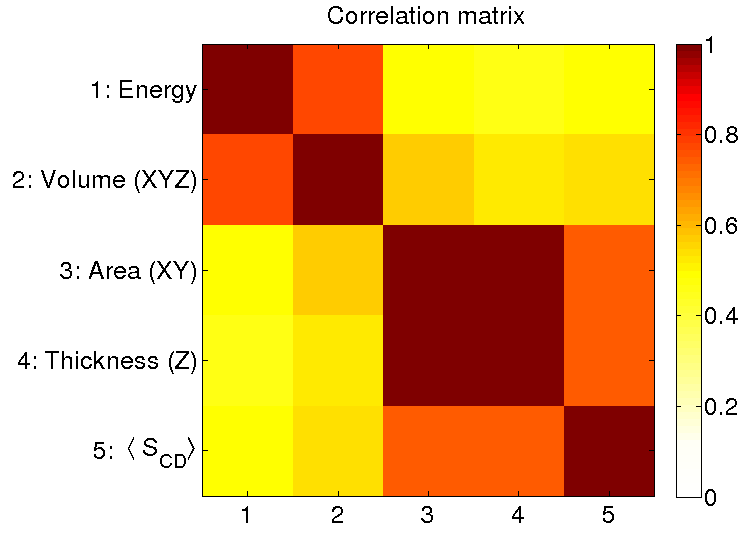}
\end{center}

\caption{Correlations in the slow thermal equilibrium fluctuations of volume and energy (top) and correlation matrix for the DMPC-f membrane (bottom). 
The normalized fluctuations of volume and potential energy shown are averaged over time intervals of 0.5 nanosecond. Data are shifted and scaled such that the average value is zero and the standard deviation is unity. A significant correlation is observed and quantified by the correlation coefficient, 
$R_{\bar{U}\bar{V}}=0.77$. This strong correlation can be associated with the tail region as seen by the similarities with Figure \ref{timeFlucTails}. The bottom panel represents the absolute values of the elements of the correlation matrix of energy, volume, membrane area, thickness and the average chain order-parameter where dark red illustrates strong correlation. Membrane area, thickness and average chain order-parameter are strongly correlated, but these quantities only correlate weakly with energy and volume. Similar results are found for the other fluid membranes (except for DMPC-g and DMPS where the energy-volume correlation is only 0.47 and 0.59, respectively), see Table \ref{Table1}.}\label{flucInTime}
\end{figure}

First, we consider the fluid DMPC-f membrane shown in Figure \ref{fluidNgel}A. Instantaneous fluctuations of volume and energy do not show any significant correlation (data not shown). This is not surprising, since a significant part of the simulation box is water, and water is known not to correlate \cite{Pedersen2008PhysRev}. However, if fluctuations are averaged over time windows of 0.5 nanosecond, volume and energy correlate strongly as shown in Figure \ref{flucInTime}A. This is quantified by the correlation coefficient,
\begin{equation}
R_{\bar{U}\bar{V}}=\frac{\langle\Delta \bar{U}\Delta \bar{V} 
\rangle }{\sqrt{\langle(\Delta \bar{U})^2\rangle\langle(\Delta \bar{V})^2\rangle}}=0.77,
\label{corrcoef}
\end{equation}
where the bar here and henceforth indicates a 0.5 ns average. $R=0$ corresponds to no correlation, whereas $|R|=1$ corresponds to perfect correlation. We will refer to $R\geq0.75$ as strong correlation. For comparison, the correlation coefficient without averaging ($R_{UV}$) is 0.35.

It is an appealing idea to establish a direct connection between thermodynamic properties and microscopic structures, and several studies have focused on the ordering of the acyl chains as the important microscopic structure \cite{nagle1980,baron2006}.
Following this philosophy, one possible explanation for the observed strong correlation is that the order of the acyl chains is the single parameter controlling the fluctuations: If chains, as a result of a thermal fluctuation, become more ordered, one expects: decrease of energy, $\bar{U}$, volume, $\bar{V}$, and area, $\bar{A}$, but increase of thickness, $\bar{Z}$, and $\langle \bar{S}_{CD}\rangle_{ch}$.
We find $R_{\bar{U}\bar{A}}=0.50$ and $R_{\bar{V}\bar{A}}=0.57$. Thus, the correlation has the expected sign, but it is significantly weaker than $R_{\bar{U}\bar{V}}$. The same is the case for $\bar{Z}$ and $\langle \bar{S}_{CD}\rangle_{ch}$. Figure \ref{flucInTime}B shows the full correlation matrix. Clearly, a single parameter description is not sufficient. We need two parameters to describe the fluctuations: one parameter controlling $\bar{V}$ and $\bar{U}$ and one ``geometrical'' parameter controlling $\bar{A}$, $\bar{Z}$, and $\langle \bar{S}_{CD}\rangle_{ch}$.

Table \ref{Table1}B shows that DMPC-f, DPPC-g, DPPC-g, DMPG, DMPS, and DMPSH all have strong volume-energy correlation ($R_{\bar{U}\bar{V}}\geq0.75$). The volumes of DMPC-g and DMPS show weaker correlation with energy. To show that this finding is not due to uncertainty from random noise, we estimated the error bar as follows: Two of the simulations, DMPC-f and DPPC-f, are about three times longer than the remaining five simulations. We used these two long simulations to estimate the error bar of the calculated correlation coefficients by dividing them into three blocks of 40 ns (regarded as uncorrelated blocks) \cite{flyvbjerg1989}. The standard deviation of the correlation coefficient is 0.03 and 0.10 for DMPC-f and DPPC-f, respectively -- or about 0.07 on average. Each of the five short runs corresponds to a single block, and thus we estimate the error bar of these to be about 0.07 (within a 67\% confidence interval). For the two longer runs we use the $\sqrt{N}$ rule and estimate an error bar of 0.07/$\sqrt(3)$=0.04. Thus, the uncertainty is smaller than the spread amongst $R_{\bar{U}\bar{V}}$'s, and the weak $\bar{V}$-$\bar{U}$ correlation of the DMPC-g and DMPS membranes is genuine -- a point returned to later.

\begin{figure}
\begin{center}
\includegraphics[width=0.45\textwidth]{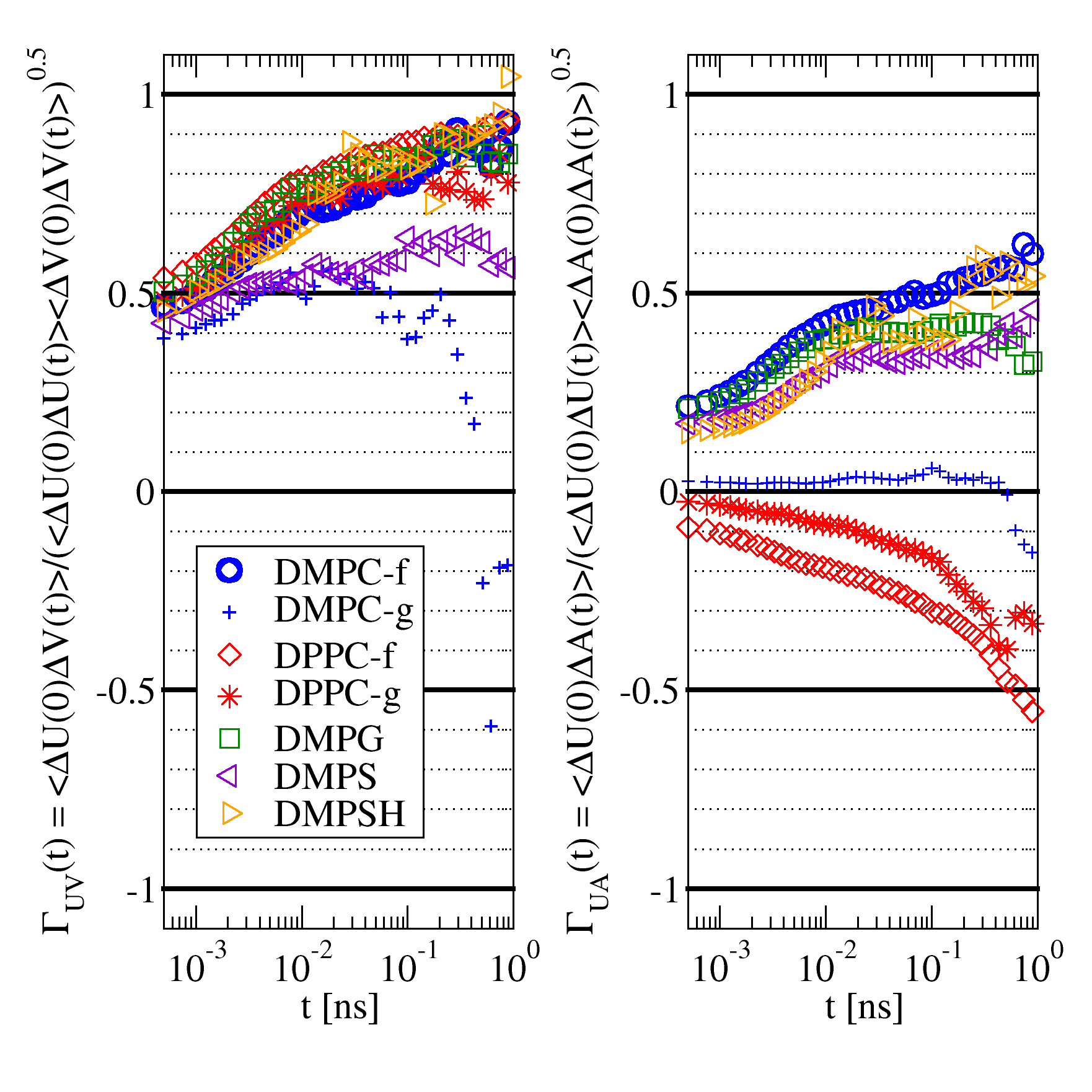}
\end{center}
\caption{Time-dependent correlation coefficient of volume-energy $\Gamma_{UV}(t)$ (left) and area-energy $\Gamma_{UA}(t)$ (right). The fast fluctuations of volume and energy, $t<10^{-2}$ ns, only correlate weakly, whereas the slow fluctuations, $t\simeq$ 1 ns, correlate strongly. Area and energy fluctuations are only weakly correlated in the investigated time regime.}\label{timecoor}
\end{figure}

The term ``slow fluctuations'' has so far been defined via the 0.5 ns averaging time window (indicated with a bar). A more general approach is to investigate the following time-dependent correlation coefficient

\begin{equation}
\Gamma_{UV}(t)=\frac{\langle\Delta U(0)\Delta V(t) \rangle }{\sqrt{\langle\Delta 
U(0)\Delta U(t)\rangle\langle\Delta V(0)\Delta V(t)\rangle}}\,.
\label{Gamma}
\end{equation}
Similarly, one defines the time-dependent energy-area correlation coefficient $\Gamma_{UA}(t)$.
Figure \ref{timecoor} shows $\Gamma_{UV}(t)$ and $\Gamma_{UA}(t)$ for all seven investigated system. In contrast to $\Gamma_{UA}(t)$, we observe strong correlation on long time scales for $\Gamma_{UV}(t)$ ($\Gamma_{UV}(t)\geq0.75$) for the five membranes where slow fluctuations of $V$ and $U$ correlate strongly consistent with $R_{\bar{U}\bar{V}}\geq0.75$ (Table \ref{Table1}B). In the following section, the slow parts of the volume and energy fluctuations are investigated via the autocorrelation functions $\langle\Delta V(0)\Delta V(t)\rangle$ and $\langle\Delta U(0)\Delta U(t)\rangle$.

\subsection{Locating the slow volume and energy fluctuations}

\begin{figure*}
\begin{center}
\includegraphics[width=0.80\textwidth]{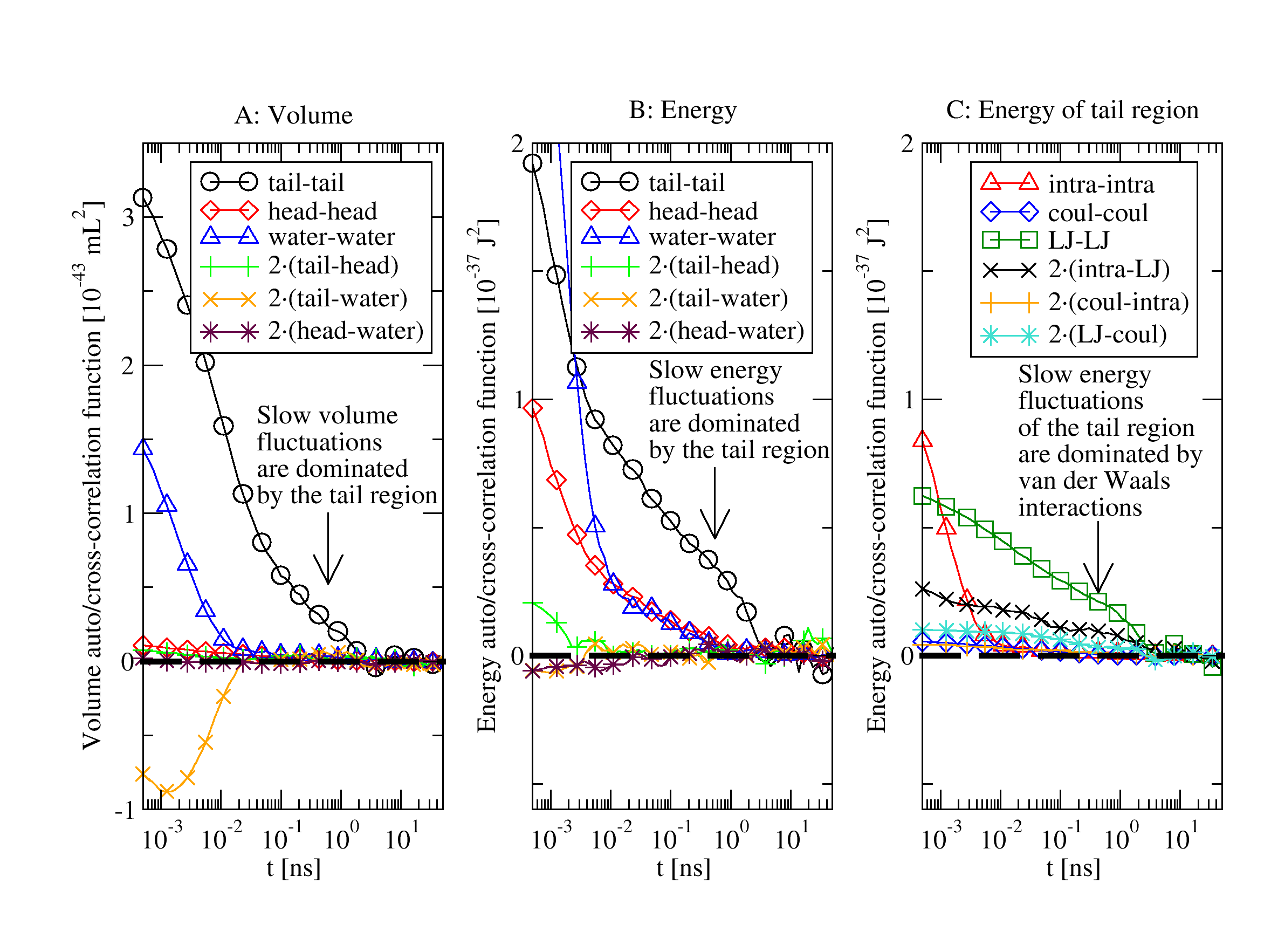}
\end{center}
\caption{Time-dependent auto- and cross-correlation function of volume and energy terms of the DMPC-f membrane. Panel A: Volume correlations functions of the tail-, head- and water regions (Equation \ref{volSplit}). Slow volume fluctuations are dominated by the tail region. The shot-time tail-water anticorrelation is probably an artifact of the way the Voronoi polyhedron is constructed (related to the tail-water interface, see text for details). Panel B: Correlation functions of energies of the three regions (Equation \ref{enerSplit}). Again, slow fluctuations are dominated by the tail region. Panel C: Energy correlations split into intramolecular, Lennard-Jones (LJ) and Coulombic energies (Equation \ref{eqx} with $x=$tail).}\label{flucOfRegions}
\end{figure*}

A membrane is a highly heterogeneous system \cite{heimburg2007book}, and it is reasonable to divide it into regions \cite{baron2006}. In the following three regions are defined: $t$, $h$, and $w$, where $t$ (tail) refers to the hydrophobic acyl-chain atoms (i.e., atoms of methylene- and methyl groups in the acyl chain), $h$ (head) refers to the hydrophilic lipid atoms (the remaining of the lipid atoms), and $w$ refers to the water atoms (and counter ions).
To identify the origin of the slow volume fluctuations, we construct Voronoi polyhedra \cite{Voronoi1908} of heavy atoms (i.e., ignoring hydrogen) and sum the Voronoi volumes for the regions $t$, $h$ and $w$. In this way, the total volume of the simulation box is divided into three terms,
\begin{equation}
V=V_t+V_h+V_w\,.
\end{equation}
The auto-correlation function of the volume in Equation \ref{Gamma} can now be split into a sum of three auto- and three cross-correlation functions,
\begin{eqnarray}\label{regions}
 \langle\Delta V(0)\Delta V(t)\rangle
&=&\langle\Delta V_t(0)\Delta V_t(t)\rangle \nonumber \\
&+&\langle\Delta V_h(0)\Delta V_h(t)\rangle \nonumber \\
&+&\langle\Delta V_w(0)\Delta V_w(t)\rangle \nonumber \\
&+&2\langle\Delta V_t(0)\Delta V_h(t)\rangle \nonumber \\
&+&2\langle\Delta V_t(0)\Delta V_w(t)\rangle \nonumber \\
&+&2\langle\Delta V_h(0)\Delta V_w(t)\rangle.
\label{volSplit}
\end{eqnarray}
Figure \ref{flucOfRegions}A shows these six functions for the DMPC-f membrane. The only nonvanishing function at long times (responsible for the slow fluctuations) is the auto-correlation function of the hydrophobic (tail) part of the membrane, $\langle \Delta V_t(0)\Delta V_t(t)\rangle$. This is quantified by $R_{\bar{V}\bar{V_t}}=0.94$ being close to unity.

The short time cross correlation between water and the tail region is significant and negative. This is, however, probably a spurious consequence of the Voronoi construction between neighboring groups of different size: When a water molecule approaches the boundary of the tail region it gains an (unfair) increase of the Voronoi volume, due to methylene groups being larger than water molecules. This show up as a negative correlation since the water molecule ``steals'' volume from the methyl groups, however, only at short times ($t<0.1$ ns).

In the simulation, the potential energy of the system consists of a sum of Lennard-Jones terms, Coulomb pair energy terms, and intramolecular binding-energies:
\begin{equation}
U=U^\textrm{intra}+\frac{1}{2}\sum_\textrm{i}\sum_{\textrm{j}\neq\textrm{i}}U_{ij}^\textrm{coul}+\frac{1}{2}\sum_\textrm{i}\sum_{\textrm{j}\neq\textrm{i}}U_{ij}^\textrm{LJ}.
\end{equation}
where $U_{ij}^\textrm{LJ}=4\varepsilon_{ij}((\sigma_{ij}/r_{ij})^{12}-(\sigma_{ij}/r_{ij})^6)$ are Lennard-Jones terms (van der Waals interactions), $U_{ij}^\textrm{coul}=q_iq_j/(4\pi\epsilon_0r_{ij})$ are Columbic terms and $U^\textrm{intra}$ is a sum of the reminder terms (which are intramolecular; harmonic bonds, angles, dihedrals and improper dihedrals).
Again, we split the total potential energy into contributions from regions of tails, heads and water,
\begin{equation}\label{enerTailSplit}
U=U_t+U_h+U_w
\end{equation}
where 
\begin{equation}\label{eqx}
U_x=U_x^\textrm{intra}+\frac{1}{2}\sum_\textrm{i = x}\sum_\textrm{j = all}U_{ij}^\textrm{coul}+\frac{1}{2}\sum_\textrm{i = x}\sum_\textrm{j=  all}U_{ij}^\textrm{LJ}\,,
\end{equation}
with $x$ either tails, heads or water.

As for the volume, the auto-correlation function of the energy fluctuations in Equation \ref{Gamma} is also split into a sum of three auto- and three cross correlation functions,

\begin{eqnarray}
 \langle\Delta U(0)\Delta U(t)\rangle
&=&\langle\Delta U_t(0)\Delta U_t(t)\rangle \nonumber \\
&+&\langle\Delta U_h(0)\Delta U_h(t)\rangle \nonumber \\
&+&\langle\Delta U_w(0)\Delta U_w(t)\rangle \nonumber \\
&+&2\langle\Delta U_t(0)\Delta U_h(t)\rangle \nonumber \\
&+&2\langle\Delta U_t(0)\Delta U_w(t)\rangle \nonumber \\
&+&2\langle\Delta U_h(0)\Delta U_w(t)\rangle.
\label{enerSplit}
\end{eqnarray}

Figure \ref{flucOfRegions}B shows the six auto- and cross correlation functions of $U_t$, $U_h$ and $U_w$. Again, the slow fluctuations are dominated by the tail region. It should be noted, though, that the slow tail-tail correlation is not quite as dominating, as for the volume fluctuations. This is quantified by $R_{\bar{U}\bar{U}_t}=0.82$ (\ref{Table1}B) not being as close to unity as $R_{\bar{V}\bar{V_t}}=0.94$: thus, slow energy fluctuations of the head- $\bar{U}_h$ and water region $\bar{U}_w$ are significant (in contrast to slow volume fluctuations of the head and water region).

Figure \ref{flucOfRegions}C shows the auto- and cross-correlation functions corresponding to an additional splitting of the tail energy into ``intramolecular interactions'', ``Coulombic interactions'' and ``van der Waals interactions'' given in Equation \ref{eqx}. The van der Waals energies dominate the energy fluctuations of the tail region.

\begin{figure}
\begin{center}
\includegraphics[width=0.45\textwidth]{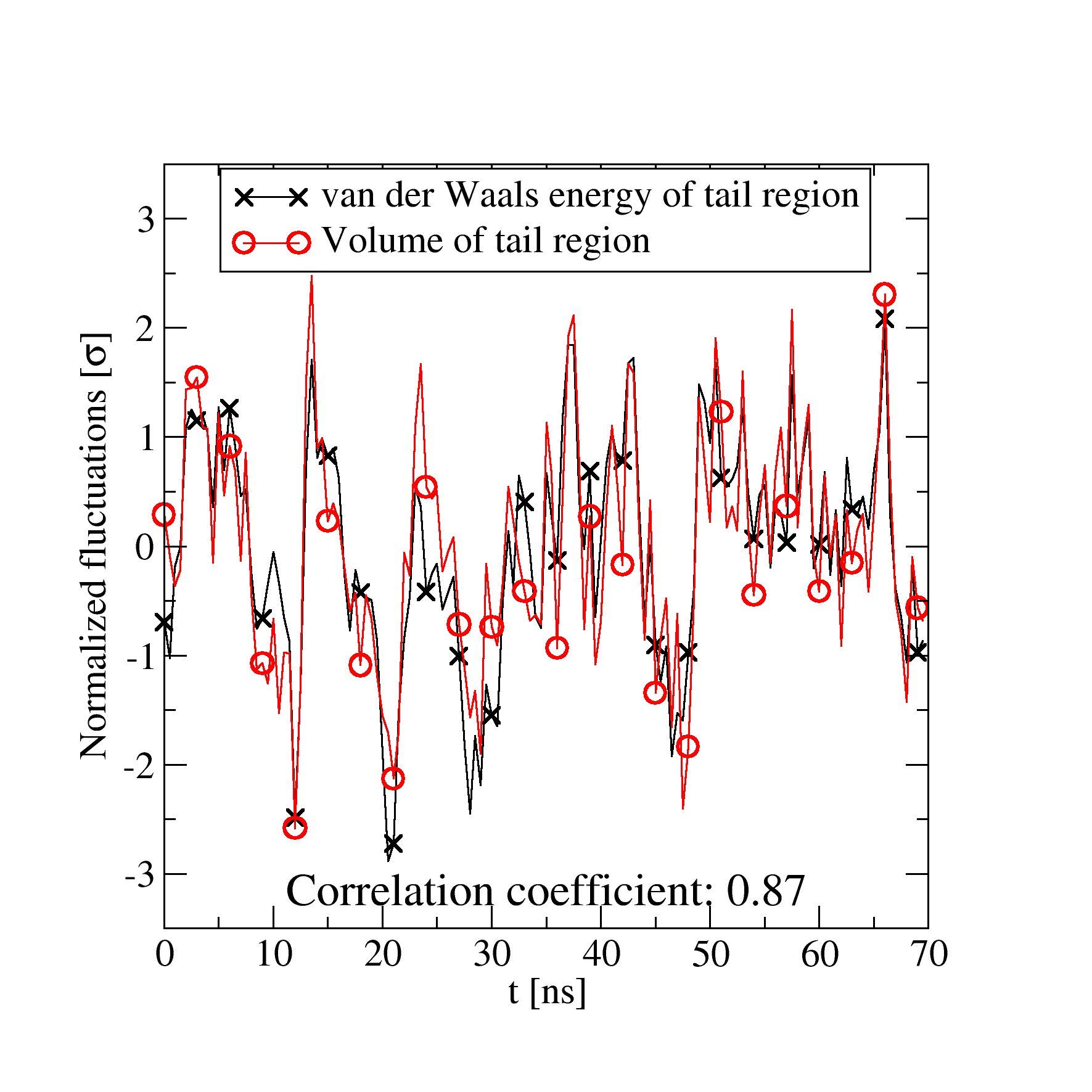}
\end{center}
\caption{Normalized fluctuations of Voronoi volume and van der Waals energy of the tail region of the DMPC-f membrane; $\bar{V}_t(t)$ and $\bar{U}_t^{LJ}(t)$. $\bar{V}_t(t)$ and $\bar{U}_t^{LJ}(t)$ are shifted and scaled so the average value is zero and the standard deviation is unity. The correlation is strong with a correlation coefficient of 0.87 (justifying Equation \ref{eqVolEnrTail}). Note the similarity with Figure \ref{flucInTime}A, thus slow $V$-$U$ fluctuations are dominated by the tail region.}\label{timeFlucTails}
\end{figure}

\section{Discussion}
In the previous section, it was demonstrated that slow energy-volume correlation (of some membranes) originates from van der Waals interactions in the tail region; thus:
\begin{equation}\label{eqVolEnrTail}
\Delta \bar{V}_t\,\propto\,\Delta \bar{U}_t^{LJ}\,,
\end{equation}
where
\begin{equation}
U_t^{LJ}=\frac{1}{2}\sum_\textrm{i=tails}\sum_\textrm{j=all}U_{ij}^\textrm{LJ}.
\end{equation}
This is consistent with our findings for simple strongly correlating liquids \cite{Bailey2008A,Bailey2008B}. It should be remembered, however, that Equation \ref{eqVolEnrTail} neglected the energy terms $U_h$ and $U_t$ with some significance. If we correlate the slow fluctuations (defined as a 0.5 ns average) of the Voronoi volume of the tail region and the van der Waals energy of the tail region, we find a correlation coefficient of $R_{\bar{V}_t\bar{U}_t^{LJ}}=$0.87, as shown in Figure \ref{timeFlucTails}. The same number for the fluctuations for whole simulation box was $R_{\bar{V}\bar{U}}=$0.77. The loss of correlation is associated with the neglected energy terms. Consistent with this, Table \ref{Table1}B shows a general correlation between $R_{\bar{V}\bar{U}}$ and $R_{\bar{U}\bar{U}_t}$, e.g. for the weakly correlated membranes, DMPS and DMPC-g, both $R_{\bar{V}\bar{U}}$ and $R_{\bar{U}\bar{U}_t}$ have low values. Apparently, slow energy fluctuations of the head region ruins the $\bar{V}$-$\bar{U}$ correlation. A better understanding of how this loss of correlation is related to various lipid head groups deserves future study.

It is convenient to define a volume-energy scaling factor via
\begin{equation}\label{gammadef}
\gamma\,=\,
\sqrt{\frac{\langle(\Delta \bar{V})^2\rangle }{\langle(\Delta \bar{U})^2\rangle}}\,,
\end{equation}
so that if $R_{\bar{U}\bar{V}}=1$ then $\Delta \bar{V}=\gamma\Delta \bar{U}$.
For the seven membranes, given in Table \ref{Table1}, we find that $\gamma$ ranges from $6.1\times10^{-4}$ mL/J (DMPC-g) to $13.2\times10^{-4}$ mL/J (DMPSH). This scaling factor is indirectly accessible though experiments: Close to the phase transition temperature curves of excess enthalpy $\Delta H(T)$ and volume $\Delta V(T)$ can be superimposed \cite{Ebel2001}. Heimburg showed that such a relation is trivial if excess volume and enthalpy of relevant micro states are proportional \cite{Heimburg1998}: $\Delta V=\gamma_{VH}\Delta H$.
The inverse $V$-$H$ scaling factor can be written as $(\gamma_{VH})^{-1}=\frac{\Delta H}{\Delta V}=\frac{\Delta U}{\Delta V}+p$. Since $p=$ 1 atm$=10^{-1}$ J/mL is much smaller than $\frac{\Delta U}{\Delta V}\simeq10^3$ J/ml (a typical value as seen in Table \ref{Table1}) then $\frac{\Delta H}{\Delta V}\simeq\frac{\Delta U}{\Delta V}$ and the $pV$ contribution to enthalpy is negligible. Moreover, the excess volume and enthalpy of the phase transition is expected to be related to the slow equilibrium fluctuations close to the transition. Thus, the scaling factor calculated from equilibrium simulations (Equation \ref{gammadef}) is closly related to the (experimental) ``phase transition'' scaling factor $\gamma_{VH}$ and we expect them to have the same value.
In agreement with this, Ebel, Grabitz and Heimburg \cite{Ebel2001} reports a scaling factor of $(9.01\pm0.71)\times10^{-4}$ mL/J and $(8.14\pm0.67)\times10^{-4}$ mL/J for large unilamellar vesicles of DMPC and DPPC, respectively -- only slightly smaller than our findings of $9.6\times10^{-4}$ mL/J and $10.2\times10^{-4}$ mL/J, respectively (Table \ref{Table1}B). This supports that we have identified the microscopic origin of the scaling of the phase transition.

We find strong energy-volume correlations in both the fluid and ordered phases. The proportionality constant $\gamma$, however, is not the same in both phases (Table \ref{Table1}B), but is about $2/3$ of the fluid value in the ordered phase. Ebel, Grabitz and Heimburg \cite{Ebel2001} reported a similar decrease (65\%; multilamellar vesicles of DMPC or DPPC) when comparing scaling factors of the pretransition with the main transition. This suggests that the configuration of the ordered membranes (DMPC-g and DPPC-g) is related to the pretransition, and that the nature of the correlations changes when passing the phase transition. Interestingly, this is in contrast to our earlier findings for crystallization of the standard Lennard-Jones liquid \cite{Bailey2008A,Bailey2008B}. More work is needed to clarify the cause of this difference between biomembranes and simple liquids.

Biologically relevant membranes are mixtures of many sorts of amphiphilic molecules and are far more complex than the membranes studied here. The origin of the strong correlations is, however, not specific, but reflects the van der Waals bonded nature of the core of the membranes. Thus, biologically relevant membranes most likely also exhibit strong correlations. How solutes in membranes effects the volume-energy correlation is an interesting subject for a future study.
Another concern could be the finite size of the simulated membranes, since fluctuations on length scales larger than the length of simulations box cannot be represented \cite{feller1996}. Again, since the origin of strong $\bar{V}$-$\bar{U}$ correlations is intrinsic to van der Waals interactions \cite{Pedersen2008PhysRev,Pedersen2008PhysRevE}, the correlations is not expected to be affected by the size.

One caveat is that although we do observe some correlation between lipid area and energy fluctuations, the correlation is weaker than the volume-energy correlation. This does not agree with the conjecture of Heimburg and coworkers \cite{Heimburg2005,Heimburg1998,Andersen2009}. Their conjecture depends on the membrane being close to the phase transition, and, it is possible that such strong $\bar{A}$-$\bar{U}$ correlation appears only when approaching the phase transition. This can be investigated by evaluating the slow $\bar{A}$-$\bar{U}$ correlation coefficient as a function of temperature. Also, it is possible that such a correlation only appears on even slower times-scales (e.g. $\mu$s).

\section{Conclusions}

This paper reports a study of thermodynamic equilibrium fluctuations of phospholipid membranes. On long time scales, we identify strong volume-energy correlations of a kind that were previously only observed for simple model liquids \cite{Pedersen2008PhysRevE,Pedersen2008PhysRev}. These correlations are mainly documented for the fluid phase, but we also show that they exist in the ordered phase. 

It may seem surprising that a complex system like a biomembrane exhibits such strong thermodynamic correlations. The identification of the origin of the correlations as deriving from the van der Waals interactions of the hydrophobic part of the membrane, however, points to a common origin of strong thermodynamic correlations in simple van der Waals liquids and biomembranes. This is consistent with the finding that there are strong energy-volume correlations in both fluid and ordered phases: The correlations do not depend on the degree of chain order, just as for simple liquids where the strong correlations survive crystallization \cite{Bailey2008A,Bailey2008B}.
We find weaker correlation between energy/volume and ``geometrical'' order parameters such as membrane area, thickness, and $S_{CD}$ (characterizing ordering of acyl-chains). Thus, one parameter is not sufficient to describe thermodynamic fluctuations.

Regarding the Heimburg-Jackson nerve signal theory, our findings largely confirm one crucial assumption of this theory, namely that volume and energy (enthalpy) correlate for microstates. We find strong correlations only on the nanosecond and longer time scales, which are, however, precisely the relevant times for nerve signals.

\section{acknowledgement}
The authors wish to thank Richard M. Venable for providing a configuration of an ordered membrane. 
NAMD was developed by the Theoretical and Computational Biophysics Group in the Beckman Institute for Advanced Science and Technology at the University of Illinois at Urbana-Champaign.
The Centre for Viscous Liquid Dynamics ``Glass and Time'' and the Centre for Biomembrane Physics ``MEMPHYS'' are both sponsored by the Danish National Research Foundation.
Simulations were performed at the Danish Center for Scientific Computing at the University of Southern Denmark.

\bibliographystyle{unsrt}
\bibliography{pedersen}

\end{document}